 \def\versionno{ twist -- version 1.0 }

\catcode`\@=11

\expandafter\ifx\csname draftcontrol\endcsname\relax\global\def\draftcontrol{0}
\fi

{\count255=\time\divide\count255 by 60
\xdef\hourmin{\number\count255}
\multiply\count255 by-60\advance\count255 by\time
\xdef\hourmin{\hourmin:\ifnum\count255<10 0\fi\the\count255}}
\def\draftdate{\number\month/\number\day/\number\year\ \ \ \hourmin }

\newcommand\makepapertitle{\par
  \begingroup
    \renewcommand\thefootnote{\@fnsymbol\c@footnote}%
    \def\@makefnmark{\rlap{\@textsuperscript{\normalfont\@thefnmark}}}%
    \long\def\@makefntext##1{\parindent 1em\noindent
            \hb@xt@1.8em{%
                \hss\@textsuperscript{\normalfont\@thefnmark}}##1}%
     \newpage
     \global\@topnum\z@   
     \@makepapertitle
     \thispagestyle{empty}\@thanks
  \endgroup
  \setcounter{footnote}{0}%
  \global\let\thanks\relax
  \global\let\makepapertitle\relax
  \global\let\@makepapertitle\relax
  \global\let\@thanks\@empty
  \global\let\@author\@empty
  \global\let\@date\@empty
  \global\let\@title\@empty
  \global\let\title\relax
  \global\let\author\relax
  \global\let\date\relax
  \global\let\and\relax
  \def\version{\let\version\@version\@gobble}
}
\def\@makepapertitle{%
  \newpage
   \ifnum\draftcontrol=1 {}
   \version\versionno
   \vskip 3em%
   \else
   \hfill\hbox to 3cm {\parbox{4cm}{\@pubnum}\hss}%
   \vskip 3em%
   \fi
   \begin{center}%
   \let \footnote \thanks
     {\LARGE \@title \par}%
     \vskip 1.5em%
     {\normalsize
       \lineskip .5em%
       \begin{center} 
         \@author
       \end{center} 
\par}%
     \vskip 1em%
     {\@bstract}%
     \end{center}%
     \vskip .5em
     \@date%
   \par
}

\gdef\@pubnum{}
\def\pubnum#1{%
  \gdef\@pubnum{#1}}

\gdef\@bstract{}
\def\Abstract#1{%
  \gdef\@bstract{%
   \parbox{\textwidth-0pc}{%
   \centerline{\bf Abstract}\penalty1000
   \noindent
   \renewcommand\baselinestretch{1.0}
   {#1}}}
}

\def\ps@paper{\let\@mkboth\@gobbletwo%
     \ifnum\draftcontrol=1
        \def\@oddfoot{\hbox to \textwidth{\tiny \versionno \hfil\tiny\draftdate}%
        \hskip -\textwidth \hbox to \textwidth{\hfil\rm\thepage\hfil}}%
     \else\def\@oddfoot{\hbox to \textwidth{\hfil\rm\thepage\hfil}}
     \fi
     \let\@evenfoot\@oddfoot
}

\def\body{\clearpage
          \pagestyle{paper}
        }
\newenvironment{acknowledgments}{%
\vskip 3.25ex
\noindent {\bf Acknowledgments}
}

\def\@version#1{\ifnum\draftcontrol=1
\typeout{}\typeout{#1}\typeout{}
\vskip3mm\centerline{\hbox{\fbox{\normalsize{\tt DRAFT -- #1 -- }
                   {\draftdate}}}}\vskip3mm
\fi}
\let\version\@version
\long\def\eqlabel#1{\ifnum\draftcontrol=1
                    \tag@false  
                    \tag*{(\theequation) \hbox to -0.2cm{\hspace{0cm}\small{#1}\hss}}
                    \refstepcounter{equation} 
                    \edef\@currentlabel{\theequation}
                    \ltx@label{#1}          
                    \else
                    \label{#1}
                    \fi
                    }
\let\st@bibitem\@bibitem
\let\st@lbibitem\@lbibitem
\ifnum\draftcontrol=1
  \def\@bibitem#1{%
    \st@bibitem{#1}\a@@label{#1}\ignorespaces}
  \def\@lbibitem[#1]#2{%
    \st@lbibitem[#1]{#2}\a@@label{#2}\ignorespaces}
  \def\a@@label#1{%
    \gdef\a@lab{\smash{\normalfont\small#1}}
    \ifvmode
      \if@inlabel
        \global\setbox\@labels\hbox{%
          \llap{\a@lab\let\a@lab\relax
                \kern\@totalleftmargin\kern\marginparsep}%
          \box\@labels}%
      \fi
    \fi}
\fi

\documentclass[12pt,letterpaper]{article}

\usepackage{amsmath,amssymb,array,calc,rotating,epsfig,psfrag}
\usepackage{hyperref}

\ifnum\draftcontrol=1
\fi

\renewcommand\baselinestretch{1.25}
\renewcommand\section{\@startsection {section}{1}{\z@}%
                                   {-3.5ex \@plus -1ex \@minus -.2ex}%
                                   {2.3ex \@plus.2ex}%
                                   {\normalfont\large\bfseries}}
\renewcommand\subsection{\@startsection{subsection}{2}{\z@}%
                                     {-3.25ex\@plus -1ex \@minus -.2ex}%
                                     {1.5ex \@plus .2ex}%
                                     {\normalfont\normalsize\bfseries}}
\renewcommand\subsubsection{\@startsection{subsubsection}{3}{\z@}%
                                     {-3.25ex\@plus -1ex \@minus -.2ex}%
                                     {1.5ex \@plus .2ex}%
                                     {\normalfont\normalsize\it}}

\numberwithin{equation}{section}

\def\projective   {{\mathbb P}}

\def\reals        {{\mathbb R}}
\def\zet          {{\mathbb Z}}

\def\revise#1       {\marginpar{\rule{2mm}{1cm} #1}}

\def\ZZ{\zet}
\def\RR{\reals}
\def\PP{\projective}
\def\RP{\RR\PP}

\def\R{{\rm R}}

\def\sqr#1#2{{\vcenter{\vbox{\hrule height.#2pt  
 \hbox{\vrule width.#2pt height#1pt \kern#1pt
 \vrule width.#2pt}\hrule height.#2pt}}}}

\def\yboxit#1#2{\vbox{\hrule height #1 \hbox{\vrule width #1
\vbox{#2}\vrule width #1 }\hrule height #1 }}
\def\fillbox#1{\hbox to #1{\vbox to #1{\vfil}\hfil}}
\def\ybox{{\lower 1.3pt \yboxit{0.4pt}{\fillbox{8pt}}\hskip-0.2pt}}

\def\comments#1{}

\def\half{{\frac12}}

\def\tr{{\rm tr}}
\def\Re{{\rm Re\hskip0.1em}}
\def\Im{{\rm Im\hskip0.1em}}

\def\ket#1{|#1\rangle}

\def\CA{{\cal A}}

\def\CD{{\cal D}}

\def\CF{{\cal F}}

\def\CH{{\cal H}}

\def\CL{{\cal L}}

\def\CN{{\cal N}}

\def\P{\BP}

\def\II{\relax{I\kern-.10em I}}

\def\IZ{\relax\ifmmode\mathchoice
{\hbox{\cmss Z\kern-.4em Z}}{\hbox{\cmss Z\kern-.4em Z}}
{\lower.9pt\hbox{\cmsss Z\kern-.4em Z}}
{\lower1.2pt\hbox{\cmsss Z\kern-.4em Z}}\else{\cmss Z\kern-.4em
Z}\fi}
\def\IB{\relax{\rm I\kern-.18em B}}
\def\IC{{\relax\hbox{$\inbar\kern-.3em{\rm C}$}}}
\def\ID{\relax{\rm I\kern-.18em D}}
\def\IE{\relax{\rm I\kern-.18em E}}
\def\IF{\relax{\rm I\kern-.18em F}}
\def\IG{\relax\hbox{$\inbar\kern-.3em{\rm G}$}}
\def\IGa{\relax\hbox{${\rm I}\kern-.18em\Gamma$}}
\def\IH{\relax{\rm I\kern-.18em H}}
\def\II{\relax{\rm I\kern-.18em I}}
\def\IK{\relax{\rm I\kern-.18em K}}
\def\IP{\relax{\rm I\kern-.18em P}}

\def\inbar{\,\vrule height1.5ex width.4pt depth0pt}

\font\cmss=cmss10 \font\cmsss=cmss10 at 7pt
\def\IR{\relax{\rm I\kern-.18em R}}

\def\BR{\IR}
\def\BP{\IP}
\def\BR{\IR}

\def\Bid{{\mathchoice {\rm {1\mskip-4.5mu l}} {\rm
{1\mskip-4.5mu l}} {\rm {1\mskip-3.8mu l}} {\rm {1\mskip-4.3mu l}}}}

\def\lp10{l_P^{10}}
\def\lp11{l_P^{11}}

\newcommand{\nc}{\newcommand}
\nc{\rnc}{\renewcommand}
\nc{\CY}{Calabi-Yau}
\nc{\CYM}{Calabi-Yau manifold}
\nc{\CYMs}{Calabi-Yau manifolds}
\nc{\DB}{D-Brane}
\nc{\DBs}{D-Branes}
\nc{\SUSY}{supersymmetry}
\nc{\Kah}{K\"ahler}
\nc{\cs}{complex structure}
\nc{\beq}{\begin{equation}}
\nc{\eeq}{\end{equation}}
\nc{\beqa}{\begin{eqnarray}}
\nc{\eeqa}{\end{eqnarray}}
\nc{\ntwo}{${\cal N}=2$}
\nc{\nOne}{${\cal N}=1$}
\nc{\hs}{\hspace{0.2in}}
\nc{\Z}{{\mathbb Z}}
\rnc{\P}{{\mathbb P}}
\rnc{\RP}{{\mathbb {RP}}}
\nc{\WP}{\mathbb{WP}}
\nc{\slag}{special Lagrangian}
\nc{\cn}{\C^n}
\nc{\rn}{\R^n}
\def\ket#1{|#1\rangle}

\nc{\SO}{SO}
\nc{\Sp}{Sp}
\nc{\SU}{SU}

\nc{\Wtree}{W_{\mathrm tree}}
\nc{\Weff}{W_{\mathrm eff}}

\catcode`\@=12

\begin{document}

\title{\Large \bf Counting chiral primaries in $\CN=1$ $d=4$ 
superconformal field theories}

\pubnum{%
hep-th/0510060}
\date{October 2005}

\author{Christian R\"omelsberger\\[0.4cm]
\it Perimeter Institute \\
\it 31 Caroline St. N. \\
\it Waterloo, ON N2L 2Y5 , Canada \\[0.2cm]
}

\Abstract{I derive a procedure to count chiral primary states in $\CN=1$ 
superconformal field theories in four dimensions. The chiral primaries are 
counted by putting the $\CN=1$ field theory on $S^3\times\BR$.
I also define an index that counts semi-short multiplets of the 
superconformal theory.
I construct $\CN=1$ 
supersymmetric Lagrangians on $S^3\times\BR$ for theories which are 
believed to flow to a conformal fixed point in the IR. For ungauged 
theories I reduce the field theory to a supersymmetric quantum mechanics,
whereas for gauge theories I use chiral ring arguments.
I count chiral primaries for $SU(2)$ SYM with three flavors and its Seiberg
dual. Those two results agree provided a new chiral ring relation holds.}

\enlargethispage{1.5cm}

\makepapertitle

\vfill \eject 

\tableofcontents

\body

\version\versionno

\section{Introduction}

Recently there has been a lot of progress in understanding the AdS-CFT
correspondence 
\cite{Berenstein:2004kk,Lin:2004nb,Berenstein:2005aa,Gopakumar:2003ns}. 
A big part of this progress was made by restricting the gauge theory
on $S^3\times\BR$ to a BPS sector. In this BPS sector the gauge theory
only the lowest harmonics of the fields contribute to the partition
function and the gauge theory simplifies to matrix quantum mechanics.

In this paper I want to restrict my attention just to gauge theories on
$S^3\times\BR$ without looking at the gravity side. I am considering the
chiral sector of $\CN=1$ field theories. In flat space this sector is only 
partly described by quantum mechanics \cite{Dijkgraaf:2002fc}, actually
it is not quantum mechanics, but a matrix integral. On $S^3\times\BR$
there is one obvious correlation function that is calculated by doing 
a restriction to quantum mechanics. This is the index\footnote{In the course
of this work related work has been published \cite{Lin:2005nh}. Part 
of the results of this work was presented in \cite{dukeseminar}.}
\beq
{\rm ind}(e^{-\beta})=\tr(-1)^Fe^{-\frac{3\beta}{2}(H-\half J)}.
\eeq
This index counts semi-short multiplets of the superconformal 
algebra\footnote{I am thankful to Juan Maldacena, Shiraz Minwalla 
and Suvrat Raju for pointing out that this index actually does not count 
chiral primaries as opposed to the claim which was made in the original 
version of this paper.}. 
A similar index is the Witten index \cite{Witten:1982df}, which is calculated
for supersymmetric field theories on $T^3\times\BR$. The Witten index 
counts supersymmetric vacua in flat space as opposed to BPS states.

In this paper, however, I am trying to develop a machinery for counting
chiral primaries, which are a subset of the semi-short multiplets. The 
chiral primaries saturate a second BPS bound. There doesn't seem to be an 
index counting those states.


It is in general hard to write down Lagrangians for $\CN=1$ superconformal
field theories. Typically those theories are defined through some field
theory in the UV which is believed to flow to a superconformal fixed point 
in the IR. The matter content and the superpotential in the UV define the
universality class of the theory. The chiral primaries of such theories
are often written down in terms of gauge invariant chiral operators 
made out of the lowest components of the chiral superfields in the Lagrangian.

The index is a topological quantity and can be calculated in a 
weak coupling regime in the UV using the Born-Oppenheimer approximation
\cite{Romelsbergerwip}. Similarly one can use the Born-Oppenheimer 
approximation to count the chiral primaries.
In this approximation method all chiral primaries actually correspond to 
gauge invariant chiral operators made out of the lowest components of the 
chiral superfields in the Lagrangian. Under the renormalization group
flow some of the chiral primaries could disappear or new chiral primaries 
could appear. Supersymmetry, however, should somewhat protect the
chiral primaries.


Typically, there are many relations in the chiral ring which can make the 
counting of chiral primaries quite involved. Using those methods one can  
compare the spectra of chiral primaries in superconformal field theories that
are related by complicated dualities like Seiberg duality, which act in very
non trivial ways on the field content.

The paper is organized as follows: In section \ref{secgroupth} I use an
$SU(2|1)$ subgroup of the $SU(2,2|1)$ superconformal algebra to derive
the BPS bound for chiral primaries and prove the existence of an index. 
In section \ref{seckilling} I construct the 
Killing spinors on $S^3\times\BR$ which are needed in section 
\ref{secfieldth} to derive the field content and the Lagrangians 
of $\CN=1$ supersymmetric field theories on this space. A similar 
approach has been used in \cite{Blau:2000xg} to construct supersymmetric 
gauge theories on curved manifolds. Furthermore I perform a twist 
on the field theory by replacing the Hamiltonian $H$
by $H-\frac{3}{2}J$, where $J$ is the R-charge. As a warmup exercise 
I count chiral primaries for some ungauged theories in section \ref{secqm}.
To describe chiral primaries in a Born-Oppenheimer approximation I do a 
consistent truncation of the field theory
to some supersymmetric quantum mechanics which has a mass term for
the bosons only. The free theory turns out to be described by a harmonic
oscillator in a frame which is rotating at the frequency of the oscillator.
From this picture it is clear that there is an infinite degeneracy of 
zero energy states corresponding to an infinite number of chiral primaries.
For Wess-Zumino models I use cohomology arguments to relate the chiral 
primaries to elements of the chiral ring. Even though it is believed that
most Wess-Zumino models are not flowing to interacting fixed points this 
is a good warmup exercise for section \ref{seccring} where I generalize 
this formalism to describe chiral primaries in gauge theories. For gauge 
theories this consistent truncation does not seem to work, but one can
still use the chiral ring arguments. As an example for an application
of this I count the chiral primaries for an $SU(2)$ SYM with 3 flavors and
its Seiberg dual, which is just a Wess-Zumino model. The the results on 
both sides of the duality agrees, however a new chiral ring relation in the SYM
is required to truncate the powers of the gaugino bilinears $S$ at
first order already instead of the well known perturbative relation 
\cite{Cachazo:2002ry}
\beq
S^2=0
\eeq
which truncates the powers of $S$ at second order.

This method could also be used to count BPS states that preserve 6 
supersymmetries in theories with extended superconformal symmetry (see e.g. 
\cite{Lin:2005nh}). Or in theories with adjoint or bifundamental 
matter like \cite{Leigh:1995ep,Klebanov:1998hh}. In those applications 
there exists a large $N$ limit and a gravity dual. One would then  
count BPS geometries.

\section{Group Theory}\label{secgroupth}

\subsection{The algebra}

The $\CN=1$ superconformal group in four dimensions is $SU(2,2|1)$, which 
contains $SU(2,2)\times U(1)_R$ as maximal bosonic subgroup. The 
supersymmetry generators $Q$ transform in the $4_1$ and their Hermitean 
conjugates $Q^\dagger$ in the $\bar 4_{-1}$.

We want to put a four dimensional $\CN=1$ superconformal field theory
on $\BR\times S^3$ which is the boundary of $AdS_5$ in global coordinates. 
The subgroup of
the conformal group consisting of isometries of $\BR\times S^3$ is
$U(1)\times SO(4)$. The supersymmetry generators decompose as follows
\beq
\begin{array}{ccc}
SU(2,2)\times U(1)_R & \rightarrow & SU(2)_l\times SU(2)_r\times U(1)\times U(1)_R \\
4_1 & \rightarrow & (2,1)_{\half,1}\oplus (1,2)_{-\half,1} \\
\bar 4_{-1} & \rightarrow & (2,1)_{-\half,-1}\oplus (1,2)_{\half,-1}
\end{array}
\eeq

Let $H$, $J_i$ and $\tilde J_i$ be the Hermitean generators of the 
isometry group $SU(2)_l\times SU(2)_r\times U(1)$. The off diagonal blocks 
in $SU(2,2)$ are the conformal generators\footnote{The undotted lower case 
Greek indices refer to the $SU(2)_l$ and the dotted lower case Greek 
indices refer to the $SU(2)_r$.} $K_{\alpha\dot\beta}$ and 
$(K_{\alpha\dot\beta})^\dagger$. In this basis the conformal algebra 
is given by
\beq\begin{array}{rcl}
{[H,H]}&=&0, \\
{[H,J_i]}&=&0, \\
{[H,\tilde J_i]}&=&0, \\
{[H,K_{\alpha\dot\beta}]}&=&-K_{\alpha\dot\beta}\\
{[J_i,J_j]}&=&i\epsilon_{ijk}\,J_k, \\
{[J_i,\tilde J_j]}&=&0, \\
{[J_i,K_{\alpha\dot\beta}]}&=&K_{\gamma\dot\beta}\,\sigma^i{}^\gamma{}_\alpha, \\
{[\tilde J_i,\tilde J_j]}&=&i\epsilon_{ijk}\,\tilde J_k, \\
{[\tilde J_i,K_{\alpha\dot\beta}]}&=&K_{\alpha\dot\gamma}\,\sigma^i{}^{\dot\gamma}{}_{\dot\alpha}, \\
{[K_{\alpha\dot\beta},K_{\gamma\dot\delta}]}&=&0, \\
{[K_{\alpha\dot\beta},(K_{\gamma\dot\delta})^\dagger]}&=&
\delta^\gamma_\alpha\,\delta^{\dot\delta}_{\dot\beta}\,H-
2\delta^\gamma_\alpha\,\sigma_{(P)}^i{}^{\dot\delta}{}_{\dot\beta}\,\tilde J_i-
2\sigma_{(P)}^i{}^\gamma{}_\alpha\,\delta^{\dot\delta}_{\dot\beta}\,J_i
\end{array}\eeq
where $\sigma_{(P)}^i{}^\alpha{}_\beta$ and 
$\sigma_{(P)}^i{}^{\dot\alpha}{}_{\dot\beta}$ are the Pauli matrices.

To construct the $\CN=1$ superconformal algebra, one has to add the
Hermitean generator $J$ of the $U(1)_R$ symmetry. $J$ commutes with all 
other bosonic generators. Finally the supersymmetry generators
are $Q_\alpha$ which transform in the $(2,1)_{-\half,-1}$ and
$S_{\dot\alpha}$ which transform in the $(1,2)_{-\half,1}$ together 
with their Hermitean conjugates $(Q_\alpha)^\dagger$ and 
$(S_{\dot\alpha})^\dagger$. The bosonic generators 
have the following commutation relations with the supercharges:
\beq\begin{array}{rclcrcl}
{[H,Q_\alpha]}&=&-\half Q_\alpha, & \qquad & 
{[H,S_{\dot\alpha}]}&=&-\half S_{\dot\alpha}, \\
{[J,Q_\alpha]}&=&-Q_\alpha, & & 
{[J,S_{\dot\alpha}]}&=&S_{\dot\alpha}, \\
{[J_i,Q_\alpha]}&=&Q_\beta\,\sigma_{(P)}^i{}^\beta{}_\alpha, & & 
{[J_i,S_{\dot\alpha}]}&=&0, \\
{[\tilde J_i,Q_\alpha]}&=&0, & & 
{[\tilde J_i,S_{\dot\alpha}]}&=&S_{\dot\beta}\,\sigma_{(P)}^i{}^{\dot\beta}{}_{\dot\alpha}, \\
{[K_{\alpha\dot\beta},Q_\gamma]}&=&0, & & 
{[K_{\alpha\dot\beta},S_{\dot\gamma}]}&=&0, \\
{[(K_{\alpha\dot\beta})^\dagger,Q_\gamma]}&=&\delta^\alpha_\gamma\,(S_{\dot\beta})^\dagger, & & 
{[(K_{\alpha\dot\beta})^\dagger,S_{\dot\gamma}]}&=&\delta^{\dot\beta}_{\dot\gamma}\,(Q_\alpha)^\dagger,
\end{array}\eeq
The anti-commutation relations between the supercharges are:
\beq\label{eqanticomm}\begin{array}{rcl}
\{Q_\alpha,Q_\beta\}&=&0, \\
\{S_{\dot\alpha},S_{\dot\beta}\}&=&0 \\
\{Q_\alpha,(Q_\beta)^\dagger\}&=&
\delta^\beta_\alpha\,(H-\frac{3}{2}J)-4\sigma_{(P)}^i{}^\beta{}_\alpha\,J_i, \\
\{S_{\dot\alpha},(S_{\dot\beta})^\dagger\}&=&
\delta^{\dot\beta}_{\dot\alpha}\,(H+\frac{3}{2}J)-
4\sigma_{(P)}^i{}^{\dot\beta}{}_{\dot\alpha}\,\tilde J_i, \\
\{Q_\alpha,S_{\dot\beta}\}&=&-2K_{\alpha\dot\beta}, \\
\{Q_\alpha,(S_{\dot\beta})^\dagger\}&=& 0.
\end{array}\eeq
The coefficients in those anticommutation relations can be fixed by
imposing the Jacobi identity and using (\ref{sigmafierz}).

\subsection{Chiral primaries}

There are important constraints from the unitatrity of representations 
of the superconformal group. We choose to label states by their quantum
numbers under the isometry group together with the $U(1)_R$ symmetry.
From the commutation relations of the other generators with the Hamiltonian 
$H$ it follows that $K_{\alpha\dot\beta}$, $Q_\alpha$ and $S_{\dot\beta}$
are lowering operators, whereas their Hermitean conjugates are raising
operators.

A primary operator is an operator that is annihilated by all the lowering
operators. i.e. it is a state of lowest energy in a representation of
the superconformal group. All other states in the same representation can
be reached by applying raising operators and operators in the isometry
group. A primary operator specifies a whole representation.

From the anti-commutation relations of the supercharges one can derive
very important unitarity bounds
\beq
-H\le\frac{3}{2}J\le H.
\eeq
Let us now concentrate on states $\ket{\psi}$ that saturate the right 
inequality. Those states are called chiral primary states. It is not 
hard to derive from the anti-commutation relations (\ref{eqanticomm}) 
that $\ket{\psi}$ satisfies
\beq
Q_\alpha\ket{\psi}=(Q_\alpha)^\dagger\ket{\psi}=J_i\ket{\psi}=0.
\eeq
Furthermore the state $S_{\dot\alpha}\ket{\psi}$ violates the unitarity 
bound and has to vanish. For this reason we get
\beq
S_{\dot\alpha}\ket{\psi}=K_{\alpha\dot\beta}\ket{\psi}=0.
\eeq
This implies that $\ket{\psi}$ is a primary operator which is also
annihilated by the raising operators $(Q_\alpha)^\dagger$.

Finally it is not hard to derive a bound on the $\tilde J_i$ quantum 
numbers which implies that $\ket{\psi}$ can have maximally spin $H$
under the $SU(2)_r$. Since there are typically infinitely many 
chiral primaries in a superconformal field theory it is useful
to count them weighted by their conformal weight
\beq
Z_\chi(x)=\tr_\chi x^H,
\eeq
where $x=e^{-\beta}$.

\subsection{An index}\label{secindex}

The superconformal algebra has a $\ZZ_2$ outer automorphism $(-1)^F$
which leaves all the bosonic symmetry generators invariant, 
negates all the supercharges and maps unitary representations to 
themselves. This suggests the existence of an index
$\tr(-1)^F$. However, since the number of states is infinite, we
need to introduce a regulator. In order for exact cancellations 
between Bosons and Fermions to happen, the regulator has to commute 
with some of the supercharges. There are two such regulators
\beq
e^{-\frac{3\beta}{2}(H-\half J)}\qquad{\rm and}\qquad 
e^{-\frac{3\beta}{2}(H+\half J)}.
\eeq
Because of the BPS bound it is easy to see that
\beq
\frac{3}{2}\left(H\pm\half J\right)\ge H
\eeq
which shows that the two regulators are good regulators for $\beta>0$.

Let us concentrate on the first regulator. It commutes with $H$, $J$, 
$J_i$, $\tilde J_i$, $Q_\alpha$ and $(Q_\alpha)^\dagger$. Those
generate the subgroup $SU(2|1)_l\times SU(2)_r\times U(1)\in SU(2,2|1)$.

The $SU(2|1)_l$ is generated by $H^\prime=H-\frac{3}{2}J$, $J_i$, $Q_\alpha$ 
and $(Q_\alpha)^\dagger$. Representations can be split up into
representations of $SU(2)_l\times U(1)_{h^\prime}$. Irreducible unitary 
representations can be labeled by the $SU(2)_l\times U(1)_{h^\prime}$ 
representation $(h^\prime,j_l)$ with the lowest $h^\prime$, 
they have to satisfy the BPS bound
\beq
2j_l\le h^\prime.
\eeq

There are three types of representations of $SU(2|1)_l$ (For a more 
general approach to representation theory of supergroups using Young 
tableaux see \cite{BahaBalantekin:1980pp}):
\begin{itemize}
\item Long representations satisfy $2j_l<h^\prime$. They contain three or four 
$SU(2)\times U(1)$ representations:
\beq\begin{array}{c}
(h^\prime,0)\oplus(h^\prime+1,\half)\oplus(h^\prime+2,0)\\{\rm and}\\
(h^\prime,j_l)\oplus(h^\prime+1,j_l-\half)\oplus(h^\prime+1,j_l+\half)
\oplus(h^\prime+2,j_l).
\end{array}\eeq
The index of a long representation vanishes.
\item Short representations satisfy $2j_l=h^\prime$, $h^\prime\ge 1$. 
They contain two $SU(2)\times U(1)$ representations:
\beq
(h^\prime=2j_l,j_l)\oplus(h^\prime+1=2j_l+1,j_l-\half).
\eeq
The index of a short representation is $\pm 1$. One can associate 
a semi-long multiplet of $SU(2,2|1)$ to those states \cite{Ferrara:1999ed},
this corresponds to a semiconserved or a conserved superfield. 
\item The trivial representation satisfies $h^\prime=j_l=0$, the index is
$\pm 1$. As shown in the last section this corresponds to chiral 
representations of $SU(2,2|1)$.
\end{itemize}

For a short representation one gets $h-\frac{3}{2}j=2j_l$.
For this reason 
\beq
{\rm ind}(e^{-\beta})=\tr(-1)^Fe^{-\frac{3\beta}{2}(H-\half J)}
\eeq
counts the number of semi-short representations  weighed by $h+j_l$ 
together with chiral primaries weighted by their conformal 
dimension. It is easy to see that inserting the other regulator into 
the trace would count short multiplets and anti-chiral primaries, where 
the shortening is due to a BPS bound involving $j_r$, the quantum number
under $SU(2)_r$.

Finally, the $SU(2)_r$ generators commute with the supercharges 
$Q_\alpha$ and $(Q_\alpha)^\dagger$ and one can restrict the index to 
specific representations of $SU(2)_r$\footnote{This can be done by inserting 
the quadratic Casimir $\tilde J_i\tilde J_i$ of $SU(2)_r$ into the trace.}:
\beq
{\rm ind}(e^{-\beta},j_r)=\tr_{j_r}(-1)^Fe^{-\frac{3\beta}{2}(H-\half J)}
\eeq
An implication of this for local conformal field theories is: 
The $SO(3)$ rotations on $S^3$
around a given fixed point are generated by the diagonal subgroup of 
$SU(2)_l\times SU(2)_r$. Since the chiral primaries are 
singlets under $SU(2)_l$ the $SU(2)_r$ quantum number is actually the
spin of the chiral primary state. Using the spin statistics theorem
one can argue that all chiral primaries contributing to the index have 
fermion number $(-1)^{2j_r}$.


If the theory has more internal symmetries, the generators of those 
can be inserted in the trace as well.

To define the index actually only the $SU(2|1)$ subalgebra of the full
conformal group was used\footnote{Actually only the subalgebra generated by
$H-\frac{3}{2}J-2J_3$, $Q_1$ and $(Q_1)^\dagger$, i.e. $SU(1|1)$ is needed 
to define the index.}. For this reason the index can be 
defined for theories which are not conformal. 
This allows to compute the index in a non conformal field theory in the 
UV, which flows to a conformal fixed point in the IR.

The $SU(2|1)$ is very similar to the supertranslation algebra. Indeed
the supertranslation algebra is a In\"on\"u-Wigner contraction of 
$SU(2|1)$. The bosonic generators of $SU(2|1)$ correspond to 
translations of $S^3\times\BR$ on which the theory is defined and the 
contraction is the large volume limit of the sphere.
The index that is defined above is actually very similar to the usual
Witten index \cite{Witten:1982df} which is defined on $T^3\times\BR$ 
and uses the supertranslation algebra as supersymmetry group. However, 
the Witten index is counting vacua of the theory in flat space as 
opposed to semi-short representations.

\section{Killing Spinors}\label{seckilling}

In order to calculate any of the quantities defined in the last section 
in an actual field theory we need to derive the classical Lagrangians 
for supersymmetric field theories on $S^3\times\BR$.
In a local field theory, the symmetries are generated by local currents.
Space time symmetries are generated by vector fields, whereas 
internal symmetries are generated by scalars with values in the Lie 
algebra of the internal symmetry group. Supersymmetries are generated 
by Grassmann valued spinor fields, the Killing spinors.

Because of the appearance of anomalous dimensions it is in general 
impossible to construct classical Lagrangians on $\BR\times S^3$ which
preserve the whole superconformal symmetry. However one can construct
classical Lagrangians on $\BR\times S^3$ which preserve the isometry 
group, the $U(1)_R$ symmetry and half of the supersymmetries.
For this purpose it is useful to introduce the anti-Hermitean operators 
\beq
\delta^{(H)}_{K^0}=iK^0 H,\quad \delta^{(J_i)}_{K^i}=iK^i J_i,\quad 
\delta^{(\tilde J_i)}_{\tilde K^i}=i\tilde K^i\tilde J_i,\quad 
{\rm and}\quad\delta^{(J)}_\Sigma=i\Sigma J
\eeq
which generate the preserved Bosonic symmetries
and the anti-Hermitean operators
\beq
\delta^{(Q)}_{\hat\zeta}=
i\hat\zeta^\alpha Q_\alpha+i(\hat\zeta^\alpha)^\ast (Q_\alpha)^\dagger,
\eeq
which generate the preserved supersymmetries. Note that 
$\hat\zeta^\alpha$ are Grassmann numbers.

The bosonic isometry algebra $U(1)\times SU(2)_l\times SU(2)_r$ can be 
realized by Killing vector fields on $\BR\times S^3$ 
\beq
K^0\partial_t,\qquad K^i\sigma_i^{(L)}\qquad{\rm and}\qquad 
\tilde K^i\sigma_i^{(R)},
\eeq
where the left and right derivatives are defined in appendix \ref{secinvforms}.
Those Killing vector fields act on the fields by through appropriately 
covariantized Lie derivatives. The $U(1)_R$ symmetry is realized by a constant
real scalar field on $\BR\times S^3$. It acts on the fields through
multiplication weighed by an appropriate charge.

In order to construct a supersymmetric field theory on $\BR\times S^3$, 
let us first construct the Killing spinors $\zeta$ which generate the
supersymmetries.
The commutation relations between the bosonic symmetry generators and
the supersymmetry generators 
\beqa
{[\delta^{(H)}_{K^0},\delta^{(Q)}_{\hat\zeta}]}&=&
\delta^{(Q)}_{-\frac{iK^0}{2}\hat\zeta}, \\
{[\delta^{(J_i)}_{K^i},\delta^{(Q)}_{\hat\zeta}]}&=&
\delta^{(Q)}_{iK^i\sigma^i\hat\zeta}, \\
{[\delta^{(\tilde J_i)}_{\tilde K^i},\delta^{(Q)}_{\hat\zeta}]}&=&0 \\
{[\delta^{(J)}_{\Sigma},\delta^{(Q)}_{\hat\zeta}]}&=&
\delta^{(Q)}_{-i\Sigma\hat\zeta}
\eeqa
imply a linear action of the bosonic symmetry generators on the
Grassmann parameters $\hat\zeta$. 

On the other hand the parameters $\hat\zeta$ should be mapped to the 
corresponding Killing spinors $\zeta$ by a linear map $M$. One can come 
up with a natural action of the bosonic symmetries on the Killing spinors.
The requirement that the diagram
\beq\label{comdiagram}\begin{array}{ccc}
\hat\zeta & \longrightarrow & g\cdot\hat\zeta \\
\downarrow & & \downarrow \\
\zeta & \longrightarrow & g\cdot\zeta
\end{array}\eeq
commutes implies the Killing spinor equations.

To deal with spinors on $\BR\times S^3$ we set up a vielbein
\beq
e^0=R_1\,dt\qquad{\rm and}\qquad e^i=R_2\,\sigma^i_{(R)},
\eeq
the gamma matrix conventions are defined in appendix \ref{secclifford}.
The relative normalization of the two constants $R_1$ and $R_2$ cannot be
fixed by the isometry algebra. It can be determined by conformal invariance
or as we will see in the next section through the supersymmetry algebra.

In this frame the action of the bosonic symmetries on the Killing spinors 
can be realized by
\beqa
R(\delta^{(H)}_{K^0})\,\zeta&=&K^0\partial_t\zeta, \\
R(\delta^{(J_i)}_{K^i})\,\zeta&=&K^i\left(\sigma_i^{(L)}+
\frac{1}{8}\epsilon_{ijk}\gamma^{jk}(\Bid+\gamma^5)\right)\zeta, \\
R(\delta^{(\tilde J_i)}_{\tilde K^i})\,\zeta&=&
\tilde K^i\left(\sigma_i^{(R)}+
\frac{1}{8}\epsilon_{ijk}\gamma^{jk}(\Bid-\gamma^5)\right)\zeta, \\
R(\delta^{(J)}_{\Sigma})\,\zeta&=&-i\Sigma\zeta.
\eeqa
For $\delta^{(H)}_{K^0}$ and $\delta^{(\tilde J_i)}_{\tilde K^i}$
the commutativity of the diagram (\ref{comdiagram}) implies
the Killing spinor equations
\beqa
\left(\partial_t+\frac{i}{2}\right)\zeta&=&0, \\
\left(\sigma_i^{(R)}+
\frac{1}{8}\epsilon_{ijk}\gamma^{jk}(\Bid-\gamma^5)\right)\zeta&=&0.
\eeqa
Those equations are integrable and have a 4 dimensional solution space.
In order to find an isomorphism to the Grassmann parameters $\hat\zeta$,
we need to restrict the Killing spinors to have positive chirality
\beq
\gamma^5\zeta=\zeta,
\eeq
which implies that
\beq
\sigma_i^{(R)}\zeta=\sigma_i^{(L)}\zeta=0.
\eeq

The commutativity of the diagram (\ref{comdiagram}) for 
$\delta^{(J)}_{\Sigma}$ transformations is automatically satisfied.
The action of $\delta^{(J_i)}_{K^i}$ on Killing spinors is actually 
given by
\beq
R(\delta^{(J_i)}_{K^i})\,\zeta=
\frac{K^i}{4}\epsilon_{ijk}\gamma^{jk}\zeta.
\eeq
The commutativity of the diagram (\ref{comdiagram}) then implies that the 
isomorphism $M$ is given by
\beq
\zeta=M\cdot\hat\zeta=
e^{-\frac{i}{2}t}\left(\begin{array}{c} \hat\zeta \\ 0 \end{array}\right).
\eeq
This result can also be derived by restricting the Killing spinors
on $AdS_5$ in global coordinates to the boundary as is done in 
\cite{Okuyama:2002zn}.

\section{Construction of the Field Theory}\label{secfieldth}

The first step in constructing a supersymmetric field theory is to find 
an action of the Killing spinors on fields, such that the anti-commutation
relations between the supercharges are satisfied. The relevant 
anti-commutation relation for realizing the supersymmetry transformations is
\beq\label{susyclosure}
[\delta^{(Q)}_{\hat\zeta_1},\delta^{(Q)}_{\hat\zeta_2}]=
-2\delta^{(H)}_{\Im(\hat\zeta_1^\dagger\hat\zeta_2)}+
3\delta^{(J)}_{\Im(\hat\zeta_1^\dagger\hat\zeta_2)}+
8\delta^{(J_i)}_{\Im(\hat\zeta_1^\dagger\sigma_{(P)}^i\hat\zeta_2)}.
\eeq
Because the closure of the supersymmetry algebra contains internal 
symmetries it is hard to use a superspace approach like in 
\cite{Wess:1992cp} to this problem instead I will use a component field 
approach.

\subsection{Chiral multiplets and the Wess-Zumino model}

Like in flat space, a chiral superfield consists of a complex scalar 
$\phi$, a chiral 
fermion $\psi$ and an auxiliary complex scalar $F$. They can be put  
together as a vector $(\phi,\psi,F)$. This chiral multiplet has to transform 
as a linear representation of the symmetry algebra. For the beginning
we assume that the chiral multiplet is not gauged. Then it transforms 
under the Bosonic symmetries as
\beqa
\delta^{(H)}_{K^0}(\phi,\psi,F)&=&
(K^0\partial_t\phi,K^0\partial_t\psi,K^0\partial_tF), \\
\delta^{(J_i)}_{K^i}(\phi,\psi,F)&=&
(K^i\sigma^{(L)}_i\phi,K^i(\sigma^{(L)}_i+
\frac{1}{4}\epsilon_{ijk}\gamma^{jk})\psi,K^i\sigma^{(L)}_iF), \\
\delta^{(\tilde J_i)}_{\tilde K^i}(\phi,\psi,F)&=&
(\tilde K^i\sigma^{(R)}_i\phi,\tilde K^i\sigma^{(R)}_i\psi,
\tilde K^i\sigma^{(R)}_iF), \\
\delta^{(J)}_\Sigma(\phi,\psi,F)&=&
(-iq\Sigma\phi,-i(q-1)\Sigma\psi,-i(q-2)\Sigma F).
\eeqa
In the presence of many internal $U(1)$ symmetries the conformal R-charge q 
of a chiral multiplet at the conformal fixed point can be determined using 
a-maximization \cite{Intriligator:2003jj}. The supersymmetry transformation 
has the form
\beq\begin{array}{l}
\delta^{(Q)}_{\hat\zeta}(\phi,\psi,F)=\\
\left(\tilde\zeta\psi,
\left(\partial_t+\frac{3iq}{2}\right)\phi\,\gamma^0\zeta^\odot-
2\sigma^{(L)}_i\phi\,\gamma^i\zeta^\odot+\zeta F,
-i\bar\zeta\gamma^0\left(\partial_t+i\frac{3q-5}{2}\right)\psi+
2i\bar\zeta\gamma^i\sigma^{(L)}_i\psi\right).
\end{array}\eeq
Using identities from appendix \ref{secclifford} one can show that 
those transformations close according to (\ref{susyclosure}).

It is not hard to verify that the Lagrangians
\beq\begin{array}{rcl}
\CL_0&=&\left(\partial_t-i\frac{3q-2}{2}\right)\phi^\ast
\left(\partial_t+i\frac{3iq-2}{2}\right)\phi-
4\sigma^{(L)}_i\phi^\ast\sigma^{(L)}_i\phi-\phi^\ast\phi\\
&&+i\bar\psi\gamma^0\left(\partial_t+i\frac{3q-2}{2}\right)\psi-
2i\bar\psi\gamma^i\left(\sigma^{(L)}_i+\frac{1}{8}\epsilon_{ijk}\gamma^{jk}\right)\psi+F^\ast F,\\
\CL_W&=&W^\prime(\phi)F-\half W^{\prime\prime}(\phi)\tilde\psi\psi+{\rm h.c.}
\end{array}\eeq
with
\beq
W^\prime(\phi)=c\phi^{-\frac{q-2}{q}}
\eeq
are supersymmetric. As expected the superpotential $W(\phi)$ has to
have R-charge 2. The kinetic terms in $\CL_0$ fixes the relative normalization
of the two vielbein coefficients $R_1$ and $R_2$ to be
\beq
R_1=-2R_2=R.
\eeq

\subsection{Vector multiplets}

A vector multiplet contains a gauge field $\CA_\mu$ a chiral fermion $\lambda$,
the gaugino, and a real auxiliary field $D$ all with values in the Lie algebra
of the gauge group\footnote{I am using anti Hermitean Lie algebra generators.}.
The gaugino and the auxiliary field transform in the adjoint representation 
of the gauge group. Let the gauge field be
\beq
\CA=\CA_0\,dt+\CA_i\,\sigma^i_{(R)}
\eeq
Then the gauge field strength is
\beqa
\CF&=&d\CA+\CA\wedge\CA\\
&=&dt\wedge\sigma^i_{(R)}
(\partial_t\CA_i-\sigma_i^{(L)}\CA_0+\CA_0\CA_i-\CA_i\CA_0)\\
&&+\half\sigma^i_{(R)}\wedge\sigma^j_{(R)}(\sigma_i^{(L)}\CA_j-
\sigma_j^{(L)}\CA_i+\epsilon_{ijk}\CA_k+\CA_i\CA_j-\CA_j\CA_i).
\eeqa
This allows to read off the components
\beqa
\CF_{0i}&=&\half(\partial_t\CA_i-\sigma_i^{(L)}\CA_0+\CA_0\CA_i-\CA_i\CA_0),\\
\CF_{ij}&=&\half(\sigma_i^{(L)}\CA_j-
\sigma_j^{(L)}\CA_i+\epsilon_{ijk}\CA_k+\CA_i\CA_j-\CA_j\CA_i).
\eeqa
The appropriate gauge covariant derivatives are
\beqa
\CD\CF&=&d\CF+\CA\wedge\CF-\CF\wedge\CA, \\
\CD_0\lambda&=&\partial_t\lambda+\CA_0\lambda-\lambda\CA_0, \\
\CD_i\lambda&=&\sigma_i^{(L)}\lambda+\CA_i\lambda-\lambda\CA_i, \\
\CD D&=&dD+\CA D-D\CA.
\eeqa
The Bianchi identity is
\beq
\CD\CF=0,
\eeq
which implies
\beq
\CD_i\CF_{j0}+\CD_j\CF_{0i}+\CD_0\CF_{ij}=\epsilon_{ijk}\CF_{0k}
\qquad{\rm and}\qquad
\epsilon_{ijk}\CD_i\CF_{jk}=0.
\eeq

Gauge transformations are generated by a Lie algebra valued generator 
$\Lambda$ as follows
\beqa
\delta^{(g)}_\Lambda\CA&=&d\Lambda+\CA\Lambda-\Lambda\CA, \\
\delta^{(g)}_\Lambda\CF&=&\CF\Lambda-\Lambda\CF, \\
\delta^{(g)}_\Lambda \lambda&=&\lambda\Lambda-\Lambda\lambda, \\
\delta^{(g)}_\Lambda D&=&D\Lambda-\Lambda D.
\eeqa

The vector multiplet $(\CA,\lambda,D)$ has to transform in a gauge 
covariant way under the bosonic symmetries. Looking at the expressions
for the components of the gauge field strength the natural transformations 
are
\beqa
\delta^{(H)}_{K^0}(\CA_0,\CA_i,\lambda,D)&=&
(0,2K^0\CF_{0i},K^0\CD_0\lambda,K^0\CD_0 D), \\
\delta^{(J_i)}_{K^i}(\CA_0,\CA_i,\lambda,D)&=&
(2K^j\CF_{j0},2K^j\CF_{ji},K^j(\CD_j+
\frac{1}{4}\epsilon_{jkl}\gamma^{kl})\lambda,K^i\CD_i D), \\
\delta^{(J)}_\Sigma(\CA_0,\CA_i,\lambda,D)&=&
(0,0,-i\Sigma\lambda,0).
\eeqa
Note that it is harder to write down the gauge covariant transformation 
law under $\delta^{(\tilde J_i)}_{\tilde K^i}$ in the frame we chose. 
Since we won't need this transformation law, we will not write it down. 
The supersymmetry transformations 
\beqa
\delta^{(Q)}_{\hat\zeta}\CA_0&=&2\Re(\bar\zeta\gamma_0\lambda), \\
\delta^{(Q)}_{\hat\zeta}\CA_i&=&-\Re(\bar\zeta\gamma_i\lambda), \\
\delta^{(Q)}_{\hat\zeta}\CF_{0i}&=&\CD_i\Re(\bar\lambda\gamma^0\zeta)-
\half\CD_0\Re(\bar\lambda\gamma^i\zeta), \\
\delta^{(Q)}_{\hat\zeta}\CF_{ij}&=&-\half\CD_i\Re(\bar\lambda\gamma^j\zeta)+
\half\CD_j\Re(\bar\lambda\gamma^i\zeta)-
\half\epsilon_{ijk}\Re(\bar\lambda\gamma^k\zeta), \\
\delta^{(Q)}_{\hat\zeta}\lambda&=&
-2i\CF_{0i}\gamma^{0i}\zeta+2i\CF_{ij}\gamma^{ij}\zeta+D\zeta, \\
\delta^{(Q)}_{\hat\zeta}\lambda^\odot&=&
2i\CF_{0i}\gamma^{0i}\zeta^\odot-2i\CF_{ij}\gamma^{ij}\zeta^\odot+
D\zeta^\odot, \\
\delta^{(Q)}_{\hat\zeta}D&=&
\Im\left(\bar\zeta\gamma^0\left(\CD_0-\frac{3i}{2}\right)\lambda\right)-
2\Im(\bar\zeta\gamma^i\CD_i\lambda)
\eeqa
close according to (\ref{susyclosure}).
A supersymmetric Lagrangian is
\beq
\CL_g=\frac{1}{g^2}\left(4\tr\CF_{0i}\CF_{0i}-8\tr\CF_{ij}\CF_{ij}+
i\tr\bar\lambda\gamma^0\CD_0\lambda-
2i\tr\bar\lambda\gamma^i
\left(\CD_i+\frac{1}{8}\epsilon_{ijk}\gamma^{jk}\right)\lambda-\tr D^2\right).
\eeq
This Lagrangian the covariantized flat-space Lagrangian.
The $\theta$-term
\beq
\CL_\theta=\theta\epsilon_{ijk}\CF_{0i}\CF_{jk}
\eeq
is separately supersymmetric. For an Abelian gauge group there is a 
Fayet-Ilioupoulous term
\beq
\CL_{FI}=\kappa\,\tr(D-\CA_0).
\eeq
Note that the integral of this Fayet-Ilioupoulous term is actually gauge
invariant. However, this Fayet-Ilioupoulous term modifies the Gauss law 
constraint drastically by producing a background charge which can lead to 
difficulties for the quantum theory. 

\subsection{Gauge invariant interactions}

Finally we want to couple chiral multiplets to gauge fields. In order to do 
so we need to replace all derivatives by covariant ones and introduce a 
few additional terms to cancel unwanted contributions. Gauge transformations
act on a chiral multiplet in the representation $\rho$ of the gauge group in 
the usual way
\beq
\delta^{(g)}_\Lambda(\phi,\psi,F)=
(-\Lambda^{(\rho)}\phi,-\Lambda^{(\rho)}\psi,-\Lambda^{(\rho)}F).
\eeq
The action of the Bosonic symmetry is easily generalized 
\beqa
\delta^{(H)}_{K^0}(\phi,\psi,F)&=&
(K^0\CD_0\phi,K^0\CD_0\psi,K^0\CD_0F), \\
\delta^{(J_i)}_{K^i}(\phi,\psi,F)&=&
(K^i\CD_i\phi,K^i(\CD_i+\frac{1}{4}\epsilon_{ijk}\gamma^{jk})\psi,
K^i\CD_iF), \\
\delta^{(J)}_\Sigma(\phi,\psi,F)&=&
(-iq\Sigma\phi,-i(q-1)\Sigma\psi,-i(q-2)\Sigma F),
\eeqa
where the gauge covariant derivatives are defined as follows
\beq
\CD_0\phi=\partial_t\phi+\CA_0^{(\rho)}\phi\qquad{\rm and}\qquad
\CD_i\phi=\sigma_i^{(L)}\phi+\CA_i^{(\rho)}\phi\qquad{\rm etc.}
\eeq
The supersymmetry transformations then have the form
\beqa
\delta^{(Q)}_\zeta\phi&=&\tilde\zeta\psi, \\
\delta^{(Q)}_\zeta\psi&=&
\left(\CD_0+\frac{3iq}{2}\right)\phi\gamma^0\zeta^\odot-
2\CD_i\phi\gamma^i\zeta^\odot+\zeta F,\\
\delta^{(Q)}_\zeta F &=& 
-i\bar\zeta\gamma^0\left(\CD_0+i\frac{3q-5}{2}\right)\psi+
2i\bar\zeta\gamma^i\CD_i\psi-2(\bar\zeta(\lambda^\odot)^{(\rho)})\phi
\eeqa
and close according to (\ref{susyclosure}).

The supersymmetric matter Lagrangian is
\beq\begin{array}{rcl}
\CL_0&=&\left(\CD_0-i\frac{3q-2}{2}\right)\phi^\dagger
\left(\CD_0+i\frac{3iq-2}{2}\right)\phi-
4\CD_i\phi^\dagger\CD_i\phi-\phi^\dagger\phi\\
&&+i\bar\psi\gamma^0\left(\CD_0+i\frac{3q-2}{2}\right)\psi-
2i\bar\psi\gamma^i\left(\CD_i+\frac{1}{8}\epsilon_{ijk}\gamma^{jk}\right)\psi+
F^\dagger F\\
&&+2i\phi^\dagger D^{(\rho)}\phi-
2i\phi^\dagger\tilde\lambda^{(\rho)}\psi+
2\bar\psi(\lambda^\odot)^{(\rho)}\phi,\\
\CL_W&=&W^\prime(\phi)F-\half W^{\prime\prime}(\phi)\tilde\psi\psi+{\rm h.c.},
\end{array}\eeq
where the superpotential $W$ is gauge invariant and has R-charge 2.

\subsection{A twist}\label{sectwist}

Since the chiral primaries saturate the BPS bound
\beq
H-\frac{3}{2}J\ge 0,
\eeq
they are the ground states of a twisted theory, where the 
original Hamiltonian $H$ is replaced by $H^\prime=H-\frac{3}{2}J$. 
In that twisted theory they can be treated using the Born-Oppenheimer
approximation. The twist is done by simply doing the replacement 
\beqa
&\lambda=e^{-\frac{3i}{2}t}\lambda^\prime,&\\
&\phi=e^{-\frac{3iq}{2}t}\phi^\prime,\quad
\psi=e^{-\frac{3i(q-1)}{2}t}\psi^\prime\quad{\rm and}\quad
F=e^{-\frac{3i(q-2)}{2}t}F^\prime&
\eeqa
in all expressions. From now on I will only deal with the twisted theory 
and omit the primes. The Lagrangians then have the form
\beqa
\CL_g&=&\frac{1}{g^2}\left(4\tr\CF_{0i}\CF_{0i}-8\tr\CF_{ij}\CF_{ij}\right)\\
&&+\frac{1}{g^2}\left(i\tr\bar\lambda\gamma^0\CD_0\lambda-
2i\tr\bar\lambda\gamma^i
\left(\CD_i+\frac{1}{4}\epsilon_{ijk}\gamma^{jk}\right)\lambda-
\tr D^2\right),\\
\CL_\theta&=&\theta\epsilon_{ijk}\CF_{0i}\CF_{jk},\\
\CL_{FI}&=&\kappa\,\tr(D-\CA_0),\\
\CL_0&=&(\CD_0+i)\phi^\dagger(\CD_0-i)\phi-
4\CD_i\phi^\dagger\CD_i\phi-\phi^\dagger\phi\\
&&+i\bar\psi\gamma^0(\CD_0-i)\psi-2i\bar\psi\gamma^i\CD_i\psi+
F^\dagger F\\
&&+2i\phi^\dagger D^{(\rho)}\phi-
2i\phi^\dagger\tilde\lambda^{(\rho)}\psi+
2\bar\psi(\lambda^\odot)^{(\rho)}\phi,\\
\CL_W&=&W^\prime(\phi)F-\half W^{\prime\prime}(\phi)\tilde\psi\psi.
\eeqa
Note that the spatial derivatives of the fermions are gauge covariant,
but have a different connection from the metric connection. This turns 
out to be more natural in the twisted theory, since it emphasizes the 
lowest energy harmonics.

This twisted theory has a supersymmetry algebra of the form
\beq\label{closuretwist}
[\delta^{(Q)}_{\hat\zeta_1},\delta^{(Q)}_{\hat\zeta_2}]=
-2\delta^{(H)}_{\Im(\hat\zeta_1^\dagger\hat\zeta_2)}+
8\delta^{(J_i)}_{\Im(\hat\zeta_1^\dagger\sigma_{(P)}^i\hat\zeta_2)}.
\eeq
with the time translations just being the appropriate gauge covariant
time derivative. The $U(1)_R$ symmetry does not have to be enforced 
anymore. This allows for inhomogeneous superpotentials. The twisted 
Lagrangians could be derived by superfield methods with the 
superspace being the $SU(2|1)$ group manifold.

Those Lagrangians look like Lagrangians of matter coupled to a background
$U(1)$ gauge field. However, this $U(1)$ is typically broken by the 
superpotential.

In order to calculate the index defined in section \ref{secindex} it is 
actually more convenient to twist the theory even more by defining the 
new Hamiltonian $H^{\prime\prime}=H-\frac{3}{2}J-2J_3$. 
In this twisted theory the index can be calculated using the 
Born-Oppenheimer approximation \cite{Romelsbergerwip}. Using 
$H^{\prime\prime}$ as a Hamiltonian amounts to going to a rotating 
coordinate system on $S^3\times\BR$.

\section{Ungauged theories}\label{secqm}

This section is devoted to the theory of a free chiral multiplet and
the Wess Zumino model. Usually, the Wess Zumino model is not really physical 
since it is believed to be IR free, but it is still
interesting to demonstrate techniques that lead to the chiral ring. 
Furthermore, the Seiberg dual of $SU(2)$ SYM with 3 flavors is a Wess Zumino
model with 15 chiral multiplets. I will count the chiral primaries of that 
theory in section \ref{secsutwo}.

\subsection{Reduction to quantum mechanics}

Chiral primaries are zero energy eigenstates of the Hamiltonian
$H^\prime=H-\frac{3}{2}J$. For this reason it is useful to reduce
the field theory on $S^3\times\BR$ to a supersymmetric quantum mechanics.
This is done by restricting to the lowest energy modes of the theory.

In a weak coupling limit the potential energy for the chiral bosons is
\beq
\sigma^{(L)}_i\phi^\dagger\sigma^{(L)}_i\phi+\phi^\dagger\phi.
\eeq
This is minimized for the S-wave mode of $\phi$. The fermion on the 
other hand minimizes the potential energy for the $SU(2)_L$ doublet
mode
\beq
\sigma^{(L)}_i\psi=0.
\eeq
With these restrictions the Lagrangian reduces to
\beq\begin{array}{rcl}
L_0&=&(\partial_t+i)\phi^\dagger(\partial_t-i)\phi-
\phi^\dagger\phi+i\bar\psi_0\gamma^0(\partial_t-i)\psi_0+F^\dagger F,\\
L_W&=&W^\prime(\phi)F-\half W^{\prime\prime}(\phi)\tilde\psi_0\psi_0.
\end{array}\eeq

The supercharges of the twisted field theory are doublets under
the $SU(2)_L$. For this reason the reduced theory, has to include
$SU(2)_L$ singlet fields together with their superpartners which are
$SU(2)_L$ doublets. The $SU(2)_L$ is now an internal symmetry of the
Lagrangian and the supersymmetry algebra still closes as in
equation (\ref{closuretwist}). The supersymmetry acts on the fields as
\beqa
\delta^{(Q)}_\zeta\phi&=&\tilde\zeta\psi, \\
\delta^{(Q)}_\zeta\psi&=&\partial_t\phi\,\gamma^0\zeta^\odot+\zeta F, \\
\delta^{(Q)}_\zeta F&=&-i\bar\zeta\gamma^0(\partial_t-i)\psi.
\eeqa
The time translation is just the derivative $\partial_t$ and the $SU(2)_L$
rotation acts on $\psi$ as
\beq
\delta^{(J_i)}_{K^i}\psi=\frac{1}{4}\epsilon_{ijk}\,\gamma^{jk}\,\psi.
\eeq

Note that this mechanics is some mechanics in a rotating frame. It is 
actually the mechanics that can be gotten from dimensional reduction
of a $\CN=1$, $d=4$ theory to $1$ dimension and adding a mass term
for the chiral bosons. The supersymmetry is generated by a time
dependent Killing spinor. For this reason the supersymmetry relates
states of different energies. A similar supersymmetric quantum
mechanics was found in \cite{Kim:2003rz}.

When doing such a reduction, one might be missing fermionic zero modes
and associated anomalies. In the twisted theory all fields are coupled 
to a constant background $U(1)$ gauge field in time direction. All 
fermions in chiral multiplets have charge $\half$ and the gauginos 
have charge $-\frac{3}{2}$ under that $U(1)$. Since this background
$U(1)$ gauge field is flat and the Pontrjagin class of the curvature
of $S^3\times\BR$ is vanishing, there are no net fermionic zero modes
and the reduction is valid.

\subsection{The canonical formalism}

The canonical momenta conjugate to $\phi$, $\psi$ are
\beq
p=(\partial_t+i)\phi^\dagger,\qquad
p^\dagger=(\partial_t-i),\qquad
\pi=-i\psi^\dagger
\eeq
and the canonical (anti-) commutation relations are
\beq
[\phi,p]=-i,\qquad [\phi^\dagger,p^\dagger]=-i,\qquad 
\{\psi_\alpha,(\psi_\beta)^\dagger\}=\delta^\beta_\alpha
\eeq
with all other (anti-) commutators vanishing. The unusual sign in
the canonical commutation relations is due to the unitarity requirement
of the fermionic sector combined with supersymmetry.

The Lagrangian then gets Legendre transformed to
the Hamiltonian
\beqa
H&=&(p-i\phi^\dagger)(p^\dagger+i\phi)+\psi^\dagger\psi\\
&&+|W^\prime(\phi)|^2-W^{\prime\prime}(\phi)\psi_1\psi_2+
(W^{\prime\prime}(\phi))^\dagger\psi_1^\dagger\psi_2^\dagger
\eeqa
The $U(1)_R$ symmetry generator is
\beq
J=-iqp\phi+iq\phi^\dagger p^\dagger-
\half(q-1)(2\psi_0^\dagger\psi_0-\Bid),
\eeq
the $SU(2)_L$ generators are
\beq
J_i=-\psi^\dagger\sigma^i_{(P)}\psi,
\eeq
the supersymmetry generators\footnote{Note that those generators are
almost the generators of usual supersymmetric quantum mechanics, where
the free part of the supercharge $-p\psi_\alpha$ is replaced by 
$-(p-i\phi^\dagger)\psi_\alpha$.} 
are
\beq
Q_\alpha=-(p-i\phi^\dagger)\psi_\alpha+
i(W^{\prime}(\phi))^\dagger\epsilon_{\alpha\beta}\psi_\beta^\dagger
\eeq
and the Fermion number operator is
\beq
(-1)^F=(-1)^{\psi^\dagger\psi}.
\eeq

\subsection{The free theory}

Let us first consider the theory of a free chiral multiplet.
The Hamiltonian is given by
\beq
H_0=(p-i\phi^\dagger)(p^\dagger+i\phi)+\psi^\dagger\psi.
\eeq
this can be analyzed using the creation and annihilation operators
\beq\begin{array}{lclcc}
a_1^\dagger=\frac{1}{\sqrt{2}}(p^\dagger-i\phi),&\qquad &
a_1=\frac{1}{\sqrt{2}}(p+i\phi^\dagger),&\qquad& 
[a_1,a_1^\dagger]=1,\\
a_2^\dagger=\frac{1}{\sqrt{2}}(p-i\phi^\dagger), &&
a_2=\frac{1}{\sqrt{2}}(p^\dagger+i\phi),&&
[a_2,a_2^\dagger]=1,\\
b_1^\dagger=\psi_1^\dagger, &&b_1=\psi_1,&&
\{b_1,b_1^\dagger\}=1, \\
b_2^\dagger=\psi_2^\dagger, &&b_2=\psi_2,&&
\{b_2,b_2^\dagger\}=1,
\end{array}\eeq
with all other (anti-) commutation relations vanishing. Those operators
define a Fock space with a vacuum
\beq
a_1\ket{0}=a_2\ket{0}=b_1\ket{0}=b_2\ket{0}=0
\eeq
and states
\beq
\ket{n_1,n_2,\eta_1,\eta_2}=\frac{1}{\sqrt{n_1!n_2!}}
(a_1^\dagger)^{n_1}(a_2^\dagger)^{n_2}
(b_1^\dagger)^{\eta_1}(b_2^\dagger)^{\eta_2}\ket{0}.
\eeq
The Hamiltonian can be written as
\beq
H_0=2a_2^\dagger a_2+b_1^\dagger b_1+b_2^\dagger b_2
\eeq
and the $U(1)_R$ generator can be written as
\beq
J=\frac{2}{3}(a_1^\dagger a_1-a_2^\dagger a_2)-
\frac{1}{3}(b_1^\dagger b_1+b_2^\dagger b_2).
\eeq
The Hamiltonian obviously has the ground states $\ket{n_1}_0=\ket{n_1,0,-,-}$ 
\beq
H_0\ket{n_1}_0=0\qquad{\rm and}\qquad 
J\ket{n_1}_0=\frac{2n_1}{3}\ket{n_1}_0.
\eeq

The Fermion number operator is given by
\beq
(-1)^F=(1-2b_1^\dagger b_1)(1-2b_2^\dagger b_2)
\eeq
which implies that all the ground states are bosons.
The chiral primaries are then counted by the partition function
\beq
Z_\chi(x)=\sum_{m=0}^\infty x^m=\frac{1}{1-x}.
\eeq

The above reasoning can be brought into a more familiar form by expressing
the chiral primary states in terms of chiral operators $\phi$
\beq
\ket{n_1}_0=\frac{1}{\sqrt{n_1!}}
(-\sqrt{2}i\phi)^{n_1}\ket{0}.
\eeq
Then the chiral primary states are just polynomials in $\phi$ as 
expected.

\subsection{The Wess-Zumino model and the chiral ring}

We have seen in section \ref{sectwist} that the superpotential of the
twisted theory can be deformed to an inhomogeneous one. The classical 
BPS equations then imply
\beq
\partial_t\phi=0\qquad{\rm and}\qquad W^\prime(\phi)=0.
\eeq
For this reason the classical theory has $n$ distinct supersymmetric 
vacua with a mass gap. This indicates that the theory has $n$ chiral 
primaries, all of the same fermion number similar as in \cite{Witten:1982df}. 
However, it is impossible to see the $U(1)_R$ 
charges of those vacua with this method. To determine those, let us 
stay with the homogeneous superpotential and find solutions to the 
quantum BPS equations. This will also make a very natural connection
with the chiral ring.

Chiral primaries satisfy
\beq
J_i\ket{\Omega}=0.
\eeq
Using $J_3$ implies that any chiral primary is a linear combination
of states of the form $\ket{m_1,m_2,-,-}$ and $\ket{m_1,m_2,+,+}$ in 
the Fock space, they are $SU(2)_l$ singlets. Let us actually split the 
Hilbert space into a subspace $\CH^{(+)}$ of states which are $SU(2)_l$ 
singlets and a subspace $\CH^{(-)}$ of states which are $SU(2)_l$ doublets
\beq
\CH=\CH^{(+)}\oplus\CH^{(-)}.
\eeq
The supersymmetry generators $(Q_\alpha)^\dagger$ are nilpotent operators
of grade one in the above grading. The kernel and the image of 
$(Q_1)^\dagger$ and $(Q_2)^\dagger$ inside $\CH^{(+)}$ agree.
Furthermore
\beq
H=\{Q_1,(Q_1)^\dagger\}
\eeq
inside $\CH^{(+)}$. For this reason the zero modes of the Hamiltonian are
harmonic representatives of the cohomology of $(Q_1)^\dagger$ inside 
$\CH^{(+)}$. This means that in order to count chiral primaries it is
enough to count $SU(2)_l$ invariant cohomology classes of $(Q_1)^\dagger$.

The supercharge $(Q_1)^\dagger$ can be written as
\beq
(Q_1)^\dagger=-\sqrt{2}a_2b_1^\dagger-
iW^\prime\left(\frac{a_2-a_1^\dagger}{\sqrt{2}i}\right)b_2.
\eeq
It is clear that the states $\ket{m,0,-,-}$ are $(Q_1)^\dagger$-closed,
using the result of the last section those states can be written as
\beq
\phi^m\ket{0}.
\eeq
Also, it is clear that states of the form 
\beq
W^\prime(\phi)\phi^m\ket{0}
\eeq
are $(Q_1)^\dagger$-exact. For this reason the cohomology of $(Q_1)^\dagger$
inside $\CH^{(+)}$ is described by polynomials in $\phi$ modulo 
$W^\prime(\phi)$, which is the chiral ring, as expected.

Note that the states $\phi^m\ket{0}$ are in general not the harmonic 
representatives of the cohomology classes, i.e. they are not the chiral
primaries. In order to prove that this is all of the cohomology, one
can note that number of chiral primaries gotten in this way agrees
with the classical argument. One can also use spectral sequence 
arguments like in \cite{Kachru:1993pg}.

All the chiral primaries have the fermion number $0$ and are counted by
\beq
Z_\chi(x)=\sum_{m=0}^{n-1} x^{\frac{3m}{n+1}}=
\frac{1-x^{\frac{3n}{n+1}}}{1-x^{\frac{3}{n+1}}}.
\eeq
The chiral primaries are all generated by $\phi$, which has quantum
numbers $(1,1)$ under $SU(2)_L\times SU(2)_R$. For this reason all
the chiral primaries are singlets under the $SU(2)_R$. In order to get 
chiral primaries of different spins, we will need to go to
gauge theories.

\section{Gauge theories and the chiral ring}\label{seccring}

The supersymmetric quantum mechanics of the last section was actually not 
gotten by a proper Kaluza Klein reduction, but by consistent truncation. 
Since the supercharge does not commute with the Hamiltonian, one always 
has to keep massive superpartners of massless fields in order to get a 
supersymmetric theory. For gauge theories there does not seem to be
a consistent truncation to quantum mechanics. However, we will use 
some of the technology of the last section and the chiral ring to 
find chiral primaries. The chiral ring that we are deriving here is
similar to the chiral ring in flat space 
\cite{Kutasov:1995ss,Cachazo:2002ry,Brandhuber:2003va}, however
it does not agree with it. For example the multi fermion operators 
derived in \cite{Beasley:2004ys} are not present in the chiral ring
on $S^3\times\BR$. In flat space the chiral ring is used to describe 
supersymmetric vacua \cite{Svrcek:2003az} similarly on $S^3\times\BR$ 
the chiral ring can be used to describe chiral primaries\footnote{ 
The number of elements in the chiral ring of flat space has been counted
in \cite{Douglas:2003um}.}.

\subsection{The chiral Ring}

In the last section we used the following two properties of chiral primaries:
They are invariant under the $SU(2)_L$ and they are 
elements of the $Q_1^\dagger$ cohomology. We will make use of those
properties to deal with gauge theories.

All $SU(2)_L$ invariant $Q_1^\dagger$ closed
states are also $Q_2^\dagger$ closed. Furthermore any $SU(2)_L$ invariant 
$Q_1^\dagger$ exact state is also $Q_2^\dagger$ exact. From this we 
conclude that the $SU(2)_L$ invariant subsector of the $Q_1^\dagger$ 
cohomology and the $Q_2^\dagger$ cohomology agree. For this reason we 
can restrict ourselves to the $Q_1^\dagger$ cohomology\footnote{This is 
actually a better situation than for the chiral ring in Minkowski
space.} $H^\bullet(Q_1^\dagger)$.

In the $SU(2)_L$ invariant sector of the Hilbert space the Hamiltonian 
can actually be written as
\beq
H=\{Q_1,Q_1^\dagger\}
\eeq
and zero energy eigenstates of the Hamiltonian again are in one to one 
correspondence with the $SU(2)_L$ invariant sector of the $Q_1^\dagger$ 
cohomology $H^0(Q_1^\dagger)$: For every cohomology representative 
$\ket{\Omega}_0\in H^0(Q_1^\dagger)$, $H\ket{\Omega}_0$ is $Q_1^\dagger$ 
exact, there is a harmonic representative 
$\ket{\Omega}_0\in H^0(Q_1^\dagger)$ in the cohomology
class of $\ket{\Omega}_0$ and every state that is annihilated 
by $H$ is $Q_1^\dagger$ closed.
This means that the zero energy eigenstates of $H$ can be represented by 
elements of $H^0(Q_1^\dagger)$.

The full $Q_1^\dagger$ cohomology $H^\bullet(Q_1^\dagger)$ actually 
describes the short representations of $SU(2|1)$ and the index is
the Euler character of the full cohomology \cite{Romelsbergerwip}.

The discussion above also shows the the elements of $H^0(Q_1^\dagger)$ are
invariant under space time translations, just as elements of the chiral 
ring of theories in flat space are.

In order to construct $H^0(Q_1^\dagger)$ it is useful to use the 
Born-Oppenheimer approximation. In a weakly coupled theory the zeroth order
approximation to the zero energy eigenstates of $H$ can be constructed
by restricting the field theory to the massless modes of the fields.
The gauge field on $S^3\times\BR$ has no massless mode, neither have
the fermions $\psi$. The only massless modes are the $(1,2)$ harmonic 
of the gaugino $\lambda_\alpha$ and the $(1,1)$ harmonic of the scalar 
$\phi$ and their conjugates. We will denote those lowest harmonics of 
$\lambda_\alpha$ and $\phi$ by $\lambda_\alpha$ and $\phi$. 
It is easy to see that those satisfy
\beq
\{Q_1^\dagger,\lambda_\alpha\}=0\qquad{\rm and}\qquad
[Q_1^\dagger,\phi]=0,
\eeq
even for non vanishing coupling, whereas their conjugates are not 
$Q_1^\dagger$ closed. For this reason the elements of
$H^0(Q_1^\dagger)$ can be represented by gauge invariant expressions
made out of $\lambda_\alpha$ and $\phi$ acting on the vacuum modulo 
relations which show that some of those states are $Q_1^\dagger$ 
exact. Note that those cohomology representatives are in general not 
the harmonic ones for interacting theories. 

This construction agrees with the operator state correspondence of
conformal field theories. This is that chiral primaries can be described
by gauge invariant polynomials in $\lambda_\alpha$ and $\phi$ acting 
on the unique conformal invariant vacuum.

The canonical commutation relations together with translation invariance
imply the relations
\beq
\{\lambda_\alpha,\lambda_\beta\}=0,\qquad
[\lambda_\alpha,\phi]=0\qquad{\rm and}\qquad
[\phi,\phi]=0
\eeq
in the chiral ring. Furthermore, by varying the elementary fields and their
canonical conjugates, one also gets the relations
\beq\label{creltwo}
F^\dagger=W^\prime(\phi)=0\qquad{\rm and}\qquad
\lambda_\alpha^{(R)}\phi^{(R)}=0
\eeq
in the chiral ring\footnote{The first relation is actually a classical 
relation and typically gets modified. In a regime where matter fields 
are integrated out it is replaced by the generalized Konishi anomaly
\cite{Cachazo:2002ry,Brandhuber:2003va}.}, where in the second equation 
the gaugino is acting on
a matter field in a given representation $R$. There might be further 
perturbative relations in the chiral ring, which are more model specific.
Those perturbative relations might get nonperturbative corrections. 

In the weak coupling nonperturbative corrections are due to instanton 
effects. On Euclidean
$S^3\times S^1$ one can indeed construct instanton configurations. 
The index is the partition function on $S^3\times S^1$ with periodic
boundary conditions on the fermions and no other operator insertions, 
for this reason it shouldn't receive any nonperturbative corrections.
The partition function for chiral primaries, however has a projection
operator on chiral primaries inserted and might receive instanton 
corrections.


This can be formulated in a different way. The only thing that matters 
about chiral ring relations for counting chiral primaries, is where 
the chiral ring gets truncated. This should not be affected by 
deformations of the relations due to nonperturbative effects.
For example in a pure $\CN=1$ super Yang Mills theory in flat space 
there is the perturbative chiral ring relation \cite{Cachazo:2002ry}
\beq
S^{N_c}=0,
\eeq
where $S=\tr\lambda_1\lambda_2$ is the gaugino bilinear. This gets modified 
by nonperturbative effects to
\beq
S^{N_c}=\Lambda^{3N_c}.
\eeq
In both cases the relation truncates the chiral ring at the same point.

On the other hand, the chiral primaries behave similar to vacua in flat
space. There are cases know in flat space where vacua are destabilized by 
nonperturbative effects \cite{Affleck:1983rr}. There the chiral ring 
constraints have no solution. Here, however we assume the existence of a 
conformally invariant vacuum and we create chiral primaries by acting with
the harmonic representatives of the chiral ring.

The example in the next section seems to suggest that our method of 
counting chiral primaries works fairly well.

All the arguments presented here only depend on the symmetry group $SU(2|1)$
and not on the full superconformal group. For this reason it might be 
interesting to extend this analysis to non-conformal theories on 
$S^3\times\BR$. 

\subsection{Seiberg duality for $SU(2)$ super Yang Mills theory with 3 
flavors}\label{secsutwo}

For a $SU(2)$ super Yang Mills theory with 3 flavors we can actually 
count the chiral primaries. The fundamental and the antifundamental
representations are the same which means that there are 6 fundamental
matter fields $q$ all with $U(1)_R$ charge $\frac{2}{3}$. 
Using the properties of group characters (see appendix A of 
\cite{Aharony:2003sx})
\beq
n_{singlet}=\int\limits_G[dg]\prod\limits_i\chi_{R_i}(g)
\eeq
and
\beq
\sum\limits_{n=0}^\infty x^{\frac{n}{2}}\chi_{sym^n(R)}(g)=
\exp\left(\sum\limits_{l=1}^\infty \frac{x^{\frac{l}{2}}}{l}\chi_R(g^l)\right)
\eeq
we can count the chiral primaries made of just matter fields. Their number is
given by the matrix integral
\beq
Z_{\chi,M}(x)=\int\limits_{SU(2)}[dg]
\exp\left(6\sum\limits_{l=1}^\infty \frac{x^{\frac{l}{2}}}{l}\chi_f(g^l)\right)^6.
\eeq
Going to an eigenvalue basis for this matrix integral introduces a 
measure factor
\beq
Z_{\chi,M}(x)=
\frac{2}{\pi}\int\limits_{\varphi=0}^\pi\sin^2\varphi\,d\varphi
\exp\left(12
\sum\limits_{l=1}^\infty \frac{x^{\frac{l}{2}}}{l}\cos(l\varphi)\right)^6=
\frac{2}{\pi}\int\limits_{\varphi=0}^\pi
\frac{\sin^2\varphi\,d\varphi}{(1-2\sqrt{x}\cos\varphi+x)^6}
\eeq
This integral can be solved
\beq
Z_{\chi,M}(x)=\frac{1+6x+6x^2+x^3}{(1-x)^9}.
\eeq

The Seiberg dual \cite{Seiberg:1994pq,Beasley:2004ys} has a trivial, 
gauge group. This means that it is a Wess Zumino model with dual 
'meson' matter fields 
$M^{ij}$ in the antisymmetric representation of the $SU(6)$ flavor
group. The superpotential is
\beq
W=\epsilon_{i_1\cdots i_6}M^{i_1i_2}M^{i_3i_4}M^{i_5i_6}.
\eeq
This imposes the chiral ring relations
\beq
\epsilon_{i_1\cdots i_6}M^{i_3i_4}M^{i_5i_6}=0.
\eeq

The dual 'meson' matter fields can be represented in terms of
Young tableaux as
\beq
{\begin{array}{|c|}
\hline
{}\;\;{} \\
\hline
{} \\
\hline
\end{array}}
\eeq
The $n$-th excited level is described by the symmetric product of
the dual 'meson' matter fields. The superpotential constraint is
that every term in the expansion of the symmetric product in young 
tableaux containing 4 or more boxes in one column is actually
vanishing. Furthermore the symmetric product implies that the
height of each column in a young tableaux is a multiple of 2. For
this reason the only young tableaux contributing is
\beq
sym^n\left(\:{\begin{array}{|c|}
\hline
{}\;\;{} \\
\hline
{} \\
\hline
\end{array}}\:\right)
{\rm mod}\:(dW=0)
=\underbrace{{\begin{array}{|c|c}
\hline
{}\;\;{} &\\
\hline
{} &\\
\hline
\end{array}}\cdots
{\begin{array}{c|c|}
\hline
&{}\;\;{} \\
\hline
&{} \\
\hline
\end{array}}}_n
\eeq
This representation has dimension
\beq
d(n)=\frac{(n+5)!\,(n+4)!}{5!\,4!\,(n+1)!\,n!}
\eeq
and the chiral primaries are counted by
\beq
Z_\chi(x)=\sum\limits_{n=0}^\infty d(n) x^{n}=
\frac{1+6x+6x^2+x^3}{(1-x)^9}.
\eeq
This shows that the number of chiral primaries is preserved under Seiberg 
duality as long as on the gauge theory side of the duality only matter 
fields are taken into account.

Because of the relation (\ref{creltwo}) the gauginos on the gauge theory 
side can only contribute in terms of gaugino bilinears (see 
\cite{Seiberg:2002jq}). Furthermore there is the perturbative chiral ring 
relation
\beq
S^2=0.
\eeq
This is not good enough yet. However, the R-charges and flavor symmetry 
allow for an extra chiral ring relation of the form
\beq
S\sim{\rm Pf}(q\cdot q).
\eeq
Perturbatively the right hand side is vanishing since the matrix $q\cdot q$
has only rank 2. Nonperturbatively this relation can make sense. As of now
I was not able to prove such a relation, but it seems to be important
for Seiberg duality to work.

\subsection{Seiberg duality in the conformal window}

In the conformal window $\frac{3N_c}{2}\le N_f\le 3N_c$ Seiberg duality
\cite{Seiberg:1994pq}
is the statement that two different gauge theories in the UV, an 'electric' 
theory and a 'magnetic' theory flow in the IR to the same fixed point.
The 'electric' theory is a $SU(N_c)$ SYM with $N_f$ flavors $Q_i$ and 
$\tilde Q^i$, whereas the 'magnetic' theory is a $SU(N_f-N_c)$ SYM with 
$N_f$ flavors $q_i$ and $\tilde q^i$ and $N_f^2$ singlet fields
$M_i{}^j$ with a superpotential
\beq
W=\tilde q^iM_i{}^jq_j.
\eeq

For such theories it seems to be much harder to count chiral primaries.
In the UV is fairly clear which gauge invariant expressions in terms
of chiral operators are supposed to be mapped to each other under
Seiberg duality. This map was made much more profound in the full gauge
theory in \cite{Kutasov:1995ss} for matter fields and for the gaugino
bilinear.

It is however less clear which chiral ring relations map to each other 
under Seiberg duality. Especially, when the gauginos are included, there
appears a similar problem as in the previous section: On the electric side
of the duality, there is the perturbative relation
\beq
S_e^{N_c}=0,
\eeq
whereas on the magnetic side there is a similar relation
\beq
S_m^{N_f-N_c}=0.
\eeq
Furthermore the electric gaugino bilinear $S_e$ and the magnetic gaugino
bilinear $S_m$ are proportional to each other. In order for the
chiral rings on both sides of the duality to truncate at the same point,
there have to be relations of the form
\beq
S_e^{N_f-N_c}\sim\det(\tilde q q)\qquad{\rm and}\qquad
S_m^{N_c}\sim\det(\tilde Q Q)
\eeq
or something similar. Again the right hand sides are vanishing perturbatively,
but might be nonvanishing by nonperturbative effects.

\begin{acknowledgments} 
\nopagebreak

\noindent Research at the Perimeter Institute is supported in part by funds from NSERC of Canada.

I would like to thank the Benasque Center for physics, the CERN theory 
division and the Duke Center for Geometry and Theoretical Physics
where part of the work was done.

I would like to thank Sujay Ashok, Paul Aspinwall, Itzhak Bars, David Berenstein, Alex Buchel, Freddy Cachazo, Rich Corrado, Mike Douglas, Laurent Freidel, Jaume Gomis, Nick Halmagyi, Ken Intriligator, Juan Maldacena, Shiraz Minwalla, Ronen Plesser, Maxim Pospelov, Suvrat Raju, Nati Seiberg, Nemani Suryanarayana and Nick Warner for useful discussions.
\end{acknowledgments}

\begin{appendix}

\section{Some conventions}\label{secclifford}

In this appendix I set up some conventions for spinors and list some
useful relations which are needed to prove the closure of the 
supersymmetry algebra and the invariance of Lagrangians under supersymmetry
transformations.
Pauli matrices are given by
\beq
\sigma_{(P)}^1=
\half\left(\begin{array}{cc} 0 & 1 \\ 1 & 0 \end{array}\right), \quad
\sigma_{(P)}^2=
\half\left(\begin{array}{cc} 0 & -i \\ i & 0 \end{array}\right), \quad
\sigma_{(P)}^3=
\half\left(\begin{array}{cc} 1 & 0 \\ 0 & -1 \end{array}\right).
\eeq
and satisfy the identity
\beq\label{sigmafierz}
\delta^\alpha_\beta\,\delta^\gamma_\delta=
\half\delta^\alpha_\delta\,\delta^\gamma_\beta+
2\sum_i\sigma_{(P)}^i{}^\alpha{}_\delta\,\sigma_{(P)}^i{}^\gamma{}_\beta.
\eeq

In the $(-+++)$ metric convention the gamma matrices can be chosen as
\beq
\gamma^0=\left(\begin{array}{cc} 0 & -\Bid \\ \Bid & 0 \end{array}\right), 
\quad
\gamma^i=2\left(\begin{array}{cc} 0 & \sigma^i \\ \sigma^i & 0 \end{array}\right).
\eeq
\beq
\gamma^5=i\gamma^{0123}=
\left(\begin{array}{cc} \Bid & 0 \\ 0 & -\Bid \end{array}\right)
\eeq
In those conventions the Hermitean conjugation is given by
\beq
(\gamma^\mu)^\dagger=-C\gamma^\mu C^{-1}\qquad{\rm with}\qquad 
C=\gamma^0\qquad{\rm and}\qquad
\bar\zeta=\zeta^\dagger C,
\eeq
the transpose is given by
\beq
(\gamma^\mu)^t=D\gamma^\mu D^{-1}\qquad{\rm with}\qquad 
D=\gamma^{135}\qquad{\rm and}\qquad
\tilde\zeta=\zeta^t D
\eeq
and the complex conjugate is given by
\beq
(\gamma^\mu)^\ast=-B\gamma^\mu B^{-1}\qquad{\rm with}\qquad 
B=\gamma^2\qquad{\rm and}\qquad
\zeta^\odot=B^{-1}\zeta^\ast.
\eeq
Chiral spinors satisfy the relations
\beqa
\gamma^5\zeta&=&\zeta, \\
\bar\zeta_1\gamma^0\zeta_2&=&-\hat\zeta_1^\dagger\hat\zeta_2, \\
\bar\zeta_1\gamma^i\zeta_2&=&-2\hat\zeta_1^\dagger\sigma_{(P)}^i\hat\zeta_2, \\
\tilde\zeta_2\gamma^\mu\zeta_1^\odot&=&i\bar\zeta_1\gamma^\mu\zeta_2, \\
\tilde\zeta_1\zeta_2&=&\tilde\zeta_2\zeta_1, \\
\epsilon_{ijk}\gamma^{jk}\psi&=&-2i\gamma^{0i}\psi
\eeqa
and the Fierz identities
\beqa
\zeta_1^\odot\,(\tilde\zeta_2\psi)&=&
-\frac{i}{2}(\bar\zeta_1\gamma_\nu\zeta_2)\,\gamma^\nu\psi, \\
\zeta_1\,(\bar\zeta_2\gamma^\mu\psi)&=&
-\half(\bar\zeta_2\gamma_\nu\zeta_1)\,\gamma^\nu\gamma^\mu\psi, \\
\zeta_1\,(\tilde\zeta_2\psi)&=&
-\half(\tilde\zeta_2\zeta_1)\,\psi-
\half(\tilde\zeta_2\gamma^{0i}\zeta_1)\,\gamma^{0i}\psi.
\eeqa

\section{Invariant forms and Killing vectors on $SU(2)$}\label{secinvforms}

The left invariant 1-forms on $SU(2)$ are given by
\beqa
\sigma^1_{(L)}&=&
\cos\varphi_3\,d\varphi_1+\sin\varphi_3\,\sin\varphi_1\,d\varphi_2, \\
\sigma^2_{(L)}&=&
\sin\varphi_3\,d\varphi_1-\cos\varphi_3\,\sin\varphi_1\,d\varphi_2, \\
\sigma^3_{(L)}&=&\cos\varphi_1\,d\varphi_2+d\varphi_3,
\eeqa
they satisfy
\beq
d\sigma^i_{(L)}=\half\epsilon_{ijk}\,\sigma^j_{(L)}\wedge\sigma^k_{(L)}.
\eeq
The dual vectors are the Killing vectors for right multiplication
\beqa
\sigma_1^{(R)}&=&\cos\varphi_3\,\partial_{\varphi_1}+
\frac{\sin\varphi_3}{\sin\varphi_1}\,\partial_{\varphi_2}-
\sin\varphi_3\,\cot\varphi_1\,\partial_{\varphi_3}, \\
\sigma_2^{(R)}&=&\sin\varphi_3\,\partial_{\varphi_1}-
\frac{\cos\varphi_3}{\sin\varphi_1}\,\partial_{\varphi_2}+
\cos\varphi_3\,\cot\varphi_1\,\partial_{\varphi_3}, \\
\sigma_3^{(R)}&=&\partial_{\varphi_3},
\eeqa
they satisfy
\beq
[\sigma_i^{(R)},\sigma_j^{(R)}]=-\epsilon_{ijk}\,\sigma_k^{(R)}
\qquad{\rm and}\qquad
\pounds_{\sigma_i^{(R)}}\sigma^j_{(L)}=-\epsilon_{ijk}\,\sigma^k_{(L)}
\eeq
Similarly, the right invariant 1-forms are 
\beqa
\sigma^1_{(R)}&=&
\cos\varphi_2\,d\varphi_1+\sin\varphi_2\,\sin\varphi_1\,d\varphi_3, \\
\sigma^2_{(R)}&=&
\sin\varphi_2\,d\varphi_1-\cos\varphi_2\,\sin\varphi_1\,d\varphi_3, \\
\sigma^3_{(R)}&=&\cos\varphi_1\,d\varphi_3+d\varphi_2
\eeqa
and the dual vectors to those are the Killing vectors for left multiplication
\beqa
\sigma_1^{(L)}&=&\cos\varphi_2\,\partial_{\varphi_1}+
\frac{\sin\varphi_2}{\sin\varphi_1}\,\partial_{\varphi_3}-
\sin\varphi_2\,\cot\varphi_1\,\partial_{\varphi_2}, \\
\sigma_2^{(L)}&=&\sin\varphi_2\,\partial_{\varphi_1}-
\frac{\cos\varphi_2}{\sin\varphi_1}\,\partial_{\varphi_3}+
\cos\varphi_2\,\cot\varphi_1\,\partial_{\varphi_2}, \\
\sigma_3^{(L)}&=&\partial_{\varphi_2},
\eeqa
they satisfy
\beqa
d\sigma^i_{(L)}&=&\half\epsilon_{ijk}\,\sigma^j_{(L)}\wedge\sigma^k_{(L)}, \\ 
{[}\sigma_i^{(R)},\sigma_j^{(R)}{]} &=& -\epsilon_{ijk}\sigma_k^{(R)}, \\
\pounds_{\sigma_i^{(L)}}\sigma^j_{(R)}&=&-\epsilon_{ijk}\,\sigma^k_{(R)}
\eeqa
and
\beqa
{[}\sigma_i^{(L)},\sigma_j^{(R)}{]} &=& 0, \\
\pounds_{\sigma_i^{(L)}}\sigma^j_{(L)}&=& 0, \\
\pounds_{\sigma_i^{(R)}}\sigma^j_{(R)}&=& 0.
\eeqa

\end{appendix}

\vfill
\pagebreak
 

\begin{thebibliography}{10}

\bibitem{Berenstein:2004kk}
D.~Berenstein, ``A toy model for the ads/cft correspondence,'' {\em JHEP} {\bf
  07} (2004) 018,
\href{http://www.arXiv.org/abs/hep-th/0403110}{{\tt hep-th/0403110}}.

\bibitem{Lin:2004nb}
H.~Lin, O.~Lunin, and J.~Maldacena, ``Bubbling ads space and 1/2 bps
  geometries,'' {\em JHEP} {\bf 10} (2004) 025,
\href{http://www.arXiv.org/abs/hep-th/0409174}{{\tt hep-th/0409174}}.

\bibitem{Berenstein:2005aa}
D.~Berenstein, ``Large n bps states and emergent quantum gravity,''
\href{http://www.arXiv.org/abs/hep-th/0507203}{{\tt hep-th/0507203}}.

\bibitem{Gopakumar:2003ns}
R.~Gopakumar, ``From free fields to ads,'' {\em Phys. Rev.} {\bf D70} (2004)
  025009,
\href{http://www.arXiv.org/abs/hep-th/0308184}{{\tt hep-th/0308184}}.

\bibitem{Dijkgraaf:2002fc}
R.~Dijkgraaf and C.~Vafa, ``Matrix models, topological strings, and
  supersymmetric gauge theories,'' {\em Nucl. Phys.} {\bf B644} (2002) 3--20,
\href{http://www.arXiv.org/abs/hep-th/0206255}{{\tt hep-th/0206255}}.

\bibitem{Lin:2005nh}
H.~Lin and J.~Maldacena, ``Fivebranes from gauge theory,''
\href{http://www.arXiv.org/abs/hep-th/0509235}{{\tt hep-th/0509235}}.

\bibitem{dukeseminar}
C.~R{\"o}melsberger, ``An index to count chiral primaries in n=1 d=4 scft's,''
  {\em String Theory Seminar at Duke} (September 15, 2005).

\bibitem{Witten:1982df}
E.~Witten, ``Constraints on supersymmetry breaking,'' {\em Nucl. Phys.} {\bf
  B202} (1982)
253.

\bibitem{Romelsbergerwip}
C.~R{\"o}melsberger, ``work in progress,''.

\bibitem{Blau:2000xg}
M.~Blau, ``Killing spinors and sym on curved spaces,'' {\em JHEP} {\bf 11}
  (2000) 023,
\href{http://www.arXiv.org/abs/hep-th/0005098}{{\tt hep-th/0005098}}.

\bibitem{Cachazo:2002ry}
F.~Cachazo, M.~R. Douglas, N.~Seiberg, and E.~Witten, ``Chiral rings and
  anomalies in supersymmetric gauge theory,'' {\em JHEP} {\bf 12} (2002) 071,
\href{http://www.arXiv.org/abs/hep-th/0211170}{{\tt hep-th/0211170}}.

\bibitem{Leigh:1995ep}
R.~G. Leigh and M.~J. Strassler, ``Exactly marginal operators and duality in
  four-dimensional n=1 supersymmetric gauge theory,'' {\em Nucl. Phys.} {\bf
  B447} (1995) 95--136,
\href{http://www.arXiv.org/abs/hep-th/9503121}{{\tt hep-th/9503121}}.

\bibitem{Klebanov:1998hh}
I.~R. Klebanov and E.~Witten, ``Superconformal field theory on threebranes at a
  calabi-yau singularity,'' {\em Nucl. Phys.} {\bf B536} (1998) 199--218,
\href{http://www.arXiv.org/abs/hep-th/9807080}{{\tt hep-th/9807080}}.

\bibitem{BahaBalantekin:1980pp}
A.~Baha~Balantekin and I.~Bars, ``Representations of supergroups,'' {\em J.
  Math. Phys.} {\bf 22} (1981)
1810.

\bibitem{Ferrara:1999ed}
S.~Ferrara and A.~Zaffaroni, ``Superconformal field theories, multiplet
  shortening, and the ads(5)/scft(4) correspondence,''
\href{http://www.arXiv.org/abs/hep-th/9908163}{{\tt hep-th/9908163}}.

\bibitem{Okuyama:2002zn}
K.~Okuyama, ``N = 4 sym on r x s(3) and pp-wave,'' {\em JHEP} {\bf 11} (2002)
  043,
\href{http://www.arXiv.org/abs/hep-th/0207067}{{\tt hep-th/0207067}}.

\bibitem{Wess:1992cp}
J.~Wess and J.~Bagger, ``Supersymmetry and supergravity,''. Princeton, USA:
  Univ. Pr. (1992) 259 p.

\bibitem{Intriligator:2003jj}
K.~Intriligator and B.~Wecht, ``The exact superconformal r-symmetry maximizes
  a,'' {\em Nucl. Phys.} {\bf B667} (2003) 183--200,
\href{http://www.arXiv.org/abs/hep-th/0304128}{{\tt hep-th/0304128}}.

\bibitem{Kim:2003rz}
N.-w. Kim, T.~Klose, and J.~Plefka, ``Plane-wave matrix theory from n = 4 super
  yang-mills on r x s**3,'' {\em Nucl. Phys.} {\bf B671} (2003) 359--382,
\href{http://www.arXiv.org/abs/hep-th/0306054}{{\tt hep-th/0306054}}.

\bibitem{Kachru:1993pg}
S.~Kachru and E.~Witten, ``Computing the complete massless spectrum of a
  landau- ginzburg orbifold,'' {\em Nucl. Phys.} {\bf B407} (1993) 637--666,
\href{http://www.arXiv.org/abs/hep-th/9307038}{{\tt hep-th/9307038}}.

\bibitem{Kutasov:1995ss}
D.~Kutasov, A.~Schwimmer, and N.~Seiberg, ``Chiral rings, singularity theory
  and electric-magnetic duality,'' {\em Nucl. Phys.} {\bf B459} (1996)
  455--496,
\href{http://www.arXiv.org/abs/hep-th/9510222}{{\tt hep-th/9510222}}.

\bibitem{Brandhuber:2003va}
A.~Brandhuber, H.~Ita, H.~Nieder, Y.~Oz, and C.~R{\"o}melsberger, ``Chiral
  rings, superpotentials and the vacuum structure of n = 1 supersymmetric gauge
  theories,'' {\em Adv. Theor. Math. Phys.} {\bf 7} (2003) 269--305,
\href{http://www.arXiv.org/abs/hep-th/0303001}{{\tt hep-th/0303001}}.

\bibitem{Beasley:2004ys}
C.~Beasley and E.~Witten, ``New instanton effects in supersymmetric qcd,'' {\em
  JHEP} {\bf 01} (2005) 056,
\href{http://www.arXiv.org/abs/hep-th/0409149}{{\tt hep-th/0409149}}.

\bibitem{Svrcek:2003az}
P.~Svrcek, ``Chiral rings, vacua and gaugino condensation of supersymmetric
  gauge theories,'' {\em JHEP} {\bf 08} (2004) 036,
\href{http://www.arXiv.org/abs/hep-th/0308037}{{\tt hep-th/0308037}}.

\bibitem{Douglas:2003um}
M.~R. Douglas, ``The statistics of string / m theory vacua,'' {\em JHEP} {\bf
  05} (2003) 046,
\href{http://www.arXiv.org/abs/hep-th/0303194}{{\tt hep-th/0303194}}.

\bibitem{Affleck:1983rr}
I.~Affleck, M.~Dine, and N.~Seiberg, ``Supersymmetry breaking by instantons,''
  {\em Phys. Rev. Lett.} {\bf 51} (1983)
1026.

\bibitem{Aharony:2003sx}
O.~Aharony, J.~Marsano, S.~Minwalla, K.~Papadodimas, and M.~Van~Raamsdonk,
  ``The hagedorn / deconfinement phase transition in weakly coupled large n
  gauge theories,'' {\em Adv. Theor. Math. Phys.} {\bf 8} (2004) 603--696,
\href{http://www.arXiv.org/abs/hep-th/0310285}{{\tt hep-th/0310285}}.

\bibitem{Seiberg:1994pq}
N.~Seiberg, ``Electric - magnetic duality in supersymmetric nonabelian gauge
  theories,'' {\em Nucl. Phys.} {\bf B435} (1995) 129--146,
\href{http://www.arXiv.org/abs/hep-th/9411149}{{\tt hep-th/9411149}}.

\bibitem{Seiberg:2002jq}
N.~Seiberg, ``Adding fundamental matter to 'chiral rings and anomalies in
  supersymmetric gauge theory','' {\em JHEP} {\bf 01} (2003) 061,
\href{http://www.arXiv.org/abs/hep-th/0212225}{{\tt hep-th/0212225}}.

\end{thebibliography}

\providecommand{\href}[2]{#2}\begingroup\raggedright\endgroup

\end{document}